\documentclass[12pt]{article}
\def\be{\begin{equation}} \def\ee{\end{equation}}
\def\bi{\begin{itemize}} \def\ei{\end{itemize}}
\def\bea{\begin{eqnarray}} \def\eea{\end{eqnarray}} \def\ba{\begin{array}}
\def\ea{\end{array}} \def\ben{\begin{enumerate}} \def\een{\end{enumerate}}

\newcommand{\eqn}[1]{(\ref{#1})}

\newcommand{\plb}[3]{Phys. Lett. B{\bf #1} ({#2}) {#3}}
\newcommand{\prd}[3]{Phys. Rev. D{\bf #1} ({#2}) {#3}}
\newcommand{\hepth}[1]{{\tt [arXiv:{#1} [hep-th]]}}

\def\br{\nonumber\\}

\begin{document}
{}~
\hfill\vbox{\hbox{hep-th/yymm.nnnn} \hbox{\today}}\break

\vskip 2.5cm
\centerline{\large \bf
Holographic flows to IR Lifshitz spacetimes}
\vskip .5cm

\vspace*{.5cm}

\centerline{  
Harvendra Singh
}
\vspace*{.25cm}
\centerline{ \it  Theory Division, Saha Institute of Nuclear Physics} 
\centerline{ \it  1/AF Bidhannagar, Kolkata 700064, India}
\vspace*{.25cm}
\vspace*{.25cm}

\vspace*{.5cm}

\vskip.5cm
\centerline{E-mail: h.singh (AT) saha.ac.in }

\vskip1.5cm

\centerline{\bf Abstract} \bigskip

Recently we studied `vanishing' horizon limits of `boosted' black 
D3-brane  geometry \cite{hsnr}. The type IIB solutions 
obtained by taking these special double limits were found to describe
nonrelativistic Lifshitz spacetimes at zero temperature.
In the present work we study these  
limits for TsT black-hole solutions which include $B$-field.
The new Galilean solutions describe a holographic RG flow from 
Schr\"odinger ($a=2$) spacetime in UV to a nonrelativistic universe in 
the IR.

\vfill 
\eject

\baselineskip=16.2pt


\section{Introduction}

There are commonly two types of
 non-relativistic or Galilean string backgrounds, 
with broken Lorentzian symmetries, which  are a subject 
of favorable attention currently \cite{son}-\cite{taylor}. 
The  geometries which possess  Schr\"odinger symmetries 
\cite{son,bala} are given as 
written as
\bea\label{sol1}
&&ds^2_{Sch}= \left( -
{\beta^2\over z^{2a}} (dx^{+})^2 +{-dx^{+}dx^{-}
+dx_i^2\over z^2} \right) +{dz^2\over  z^2}  
\eea 
and the
others with Lifshitz-like symmetries  \cite{Horava,kachru} are
written as 
\bea\label{sol1a}
&&ds^2_{Lif}= \left( -{\beta^2\over z^{2a}} dt^2
+{dx_i^2\over z^2} \right) +{dz^2\over  z^2}.  
\eea 
In both these cases, $x^i~(i=1,...,d)$ are flat spatial coordinates, 
$x^{\pm}$ are the light-cone coordinates, $z$ is the holographic 
direction and parameter $a$ is known as the dynamical 
exponent of the Galilean geometry. Some of
these  nonrelativistic geometries are  claimed 
  to be describing  strongly coupled scaling phenomena near
quantum critical points in  dual nonrelativistic  CFTs \cite{son,bala}.
The Schr\"odinger  AdS spacetimes with dynamical exponent $a=2$ can 
however be embedded in string theory
as it has been shown in  \cite{herzogrev,malda}. The Schr\"odinger
 spacetimes with  $a=3$ can also be 
found in the massive type IIA 
string theory \cite{hs}. 
More recently, 
various Lifshitz spacetimes with light-like deformation were
constructed as string solutions, as 
shown in \cite{bala3}. These solutions require  nontrivial dilaton 
field as well as metric deformations. Such solutions are further 
generalised in \cite{donos10}. These recent Lifshitz solutions do 
avoid  
the early `no-go' results of \cite{Li} because the field 
ans\"atze are some what 
 less restrictive. 
This implies that a wider class of
 Lifshitz-like solutions can  be found in string theory if we suitably 
excite other fields in the Lifshitz background.  
The hope is that some of 
these solutions could potentially describe  interesting scaling 
phenomena in dual field theories  where Lorentzian symmetry 
is explicitly broken and the system behaves quantum mechanically.
Some new Lifshitz-like solutions with Janus-like configurations have been 
presented in \cite{nishi}.

\subsection{A circle fibration over Lifshitz spacetime}
In a recent  work \cite{hsnr} we  discussed a new type of nonrelativistic 
$AdS_5$ geometry which had one of the lightcone 
direction namely $x^{+}$ (time) being null while  $x^{-}$ being
compact \footnote{Most of our analysis in this work 
goes through even for the noncompact cases.}
\bea\label{sol1b}
&&ds^2_{}= \left( -r^2{dx^{+}dx^{-}}+
{\beta^2\over 4r^2 } (dx^{-})^2 +r^2(dx_1^2 +dx_2^2)\right) +{dr^2\over  
r^2} 
\eea
It was obtained by taking  special vanishing horizon limits
of  `boosted' black D3-brane geometry. For the thermodynamics 
these limits correspond to taking `zero' 
temperature (or condensation) limits with vanishing chemical potential. 
The  Galilean geometry \eqn{sol1b} may also 
be seen as a Lifshitz spacetime  having a circle fibration
\bea
 &&
ds^2_{}\equiv ds^2_{Lif_4} +{\beta^2\over 4 r^2}\Xi^2
\eea
where 4-dimensional Lifshitz spacetime is 
\be
ds^2_{Lif_4}= -{r^6\over \beta^2}(dx^{+})^2
+r^2(dx_1^2 +dx_2^2)
 +{dr^2\over  r^2} , ~~~~
{\rm ~and}~~~~ \Xi= 
(dx^{-}-{2r^4\over\beta^2} dx^{+}).
\ee
The size of the $x^-$ fiber varies all over the Lifshitz base. Since $x^-$ 
is a circle this geometry obviously cannot be trusted  
at   large $r$ where fiber size shrinks, nevertheless in some finite 
region inside the 
bulk we can  trust this classical background. 
We shall discuss how to include a boundary configuration in these solutions.
 
In this paper we wish to extend our analysis to include backgrounds with 
NS-NS $B$-field. Specially we  focus on the  finite temperature 
T-s-T backgrounds obtained given in \cite{malda}. One useful feature of the TsT 
black-holes  is that they 
are by construction asymptotically Schr\"odinger geometries with dynamical 
exponent $a=2$. We would like to study what happens to these solutions under
the `vanishing' horizon double limits.
The  paper is organised as follows. In  section-II we review the 
boosted black D3-brane solution and  the vanishing horizon double limits 
where  the black hole horizon shrinks 
to zero value while the `boost' is simultaneously taken to be very large. The 
solutions thus obtained 
describe   zero temperature non-relativistic Lifshitz geometry. These 
solutions are not well defined at the boundary. We 
 discuss the issue of the boundary and propose a new gravity 
solution which includes a finite boundary configuration.
In section III we repeat our analysis for TsT black hole solutions which 
have asymptotic Schr\"odinger symmetries.
The section IV has the conclusions.

\section{Simultaneous double limits of 'boosted' black 3-branes}

\subsection{Review}
This section contains the review of our previous work \cite{hsnr}.
We are particularly interested in studying the DLCQ  of  
$AdS_5$ geometry as described in \cite{malda}.
We start with the `boosted' version of black D3-branes  \cite{malda}
where the near horizon solution is
\bea\label{sol2}
&&ds^2_{D3}=r^2\left( -{1+f\over 2}dx^{+}dx^{-}+{1-f\over 4}[
\lambda^{-2}(dx^{+})^2
 +\lambda^2 (dx^{-})^2]
+dx_1^2 +dx_2^2\right) 
+{dr^2\over f r^2}  + d\Omega_5^2 \, ,\br
&& F_5=4(1+\ast)Vol(S^5)
\eea 
where $d\Omega_5^2$ is the 
line element of a unit size five-sphere $S^5$. 
The  function $f(r)=1-r_0^4/r^4$, with 
$r=r_0$ being  the horizon location and the boundary is at $r\to\infty$. 
The overall $AdS_5$ 
radius, $L$, has been set to unity.
The black D3-branes \eqn{sol2} have large but finite momentum along 
compact direction $x^{-}$, 
$x^{-}\sim x^{-}+2\pi r^-$. (We shall represent this circle as $\tilde 
S^1$ 
through out this paper as it will be present everywhere.)
The boundary conformal field theory will have  a DLCQ description in a 
given (discrete) momentum sector.  However,
due to $x^-$ being compact, the geometry is 
well defined only in the interior region and not near the 
boundary. 
It should however be kept in mind that, since $x^-$ is compact we
simply cannot take $r_0\to 0$, as it will make 
$x^-$ a null direction.\footnote{ Note, it generally is not a 
problem when 
 $x^{-}$ is noncompact, because in that case  setting $r_0= 0$  
simply describes an extremal (BPS) limit which takes us to an
ordinary AdS spacetime whose holographic dual  is $N=4$
super-Yang-Mills theory at large 't Hooft coupling.}
Note that in \eqn{sol2} the size of $x^-$ 
circle  shrinks as we go near the boundary, but 
due to backreaction of the large lightcone momentum it stays finite  
within the bulk where we can trust this solution. 

The boost  parameter
$\lambda$ physically controls the size of $x^{-}$. 
As we see that  $r_0^4\lambda^2$ effectively 
measures the size of this 
circle, therefore we  consider
a combined limit in which the size of 
horizon is allowed to shrink  while boost is simultaneously taken to 
be large such that
\be\label{sol21}
r_0\to 0,~~~~ \lambda\to\infty, ~~~~r_0^4\lambda^2=\beta^2= {\rm fixed}.
\ee
 In which case we find \cite{hsnr}
\bea\label{sol22}
&&(1+f)\to 2+O(r_0^4),~~~~  {1-f\over\lambda^2}\to  O({r_0^4\over\lambda^2}) \br
&& {(1-f)\lambda^2}\to  {\beta^2\over r^4}
\eea
and the solution \eqn{sol2} simply reduces to 
\bea\label{sol23}
&&ds^2_{10}= r^2\left( -dx^{+}dx^{-}+{\beta^2\over 4 r^4} (dx^{-})^2
+dx_1^2 +dx_2^2\right) +{dr^2\over  r^2}  + d\Omega_5^2  \br
&& F_5=4(1+*)Vol(S^5) 
\eea 
which itself is a complete solution of type IIB string theory  at zero 
temperature. There is no curvature singularity. 
Actually the spacetime \eqn{sol23} 
is  still a direct product of
a `constant negative curvature spacetime' and a 5-sphere
 but it is a Galilean geometry
and crucially the coordinate $x^{+}$ (time) is null. Here
we must clarify our use of terminology a bit. These `Galilean' 
negative curvature  
spaces are spacetimes of constant negative curvature, $R=-20$,
for which Ricci tensor is given by $R_{\mu\nu}=-4 g_{\mu\nu}$, 
but the curvature tensor differs from being 
\bea
R_{\mu\nu\lambda\rho}\propto 
-(g_{\mu\lambda}g_{\nu\rho}-g_{\nu\lambda}g_{\mu\rho})
\nonumber \eea 
usually by  
additive constants. So strictly speaking
these Galilean spaces are not $AdS$ spaces.\footnote{Such 
constant curvature spaces 
were previously known as {\it Kaigorodov} spaces 
in four dimensions \cite{kaigo}. 
We thank the anonymous referee for bringing up this fact to our knowledge 
and for the references \cite{cvetic,kaigo} 
where generalized Kaigorodov spaces are 
studied in the context of pp-waves.}
However 
 we could always rewrite \eqn{sol23} as
\bea\label{sol24}
&&ds^2_{10}= ds^2_{Lif_4}+{\beta^2\over 4 r^2} 
\Xi^2 + d\Omega_5^2  .
\eea 
Therfore
 the Galilean geometry \eqn{sol23} is indeed a Lifshitz 
four-universe
along with a fibered  product space $\tilde S^1\times S^5$. 
It will represent a  well defined
system of  Kaluza-Klein particles if we  compactify the solution to four 
dimensions.
In particular,  the 9-dimensional type II solution schematically 
will be like
\bea\label{sollif}
&&ds^2_{9}= \left( -{r^6\over \beta^2}(dx^{+})^2+
r^2( dx_1^2 +dx_2^2) +{dr^2\over  r^2}\right)  + d\Omega_5^2 \br
&& e^{-2\phi}={\beta\over 2 r}, ~~~F_2^{(-)}\equiv dA^{(-)}= 
-{8r^3\over\beta^2}dr\wedge 
dx^{+} ,
\eea 
there will be a scalar $\sigma$  from internal metric component 
 $e^{2\sigma}=G_{--}$ which  couples to the KK 
gauge fields $A^{(-)}$. There will also be 3-form tensor field as a 
result of the reduction of the 4-form. 
That is we are effectively  dealing with system of KK fields coupled to 
scalar field and a dilaton field in 
4-dimensional Lifshitz universe while rest of the directions are all 
compact. 
 
In summary,   taking the double
   limits \eqn{sol21} of  the `boosted' black 
D3-branes allows us to exclusively  `zoom onto' the KK system in a 
Lifshitz universe such as \eqn{sollif}. 
The Lifshitz geometry \eqn{sol23} 
is inherently nonrelativistic.
There is an asymmetric scale (dilatation) invariance
\be
r\to (1/\xi) r , ~~~~
x^{-}\to \xi^{-1}  x^{-},~~~~
x^{+}\to \xi^3 x^{+},~~~x_{1,2}\to \xi x_{1,2}
\label{a3}\ee
where time scales with scaling dimension 3 and therefore the dynamical 
exponent is $3$. 
There are also invariances under constant shifts (translations)
 as well as   rotations in $x^1-x^2$ plane, see \cite{hsnr}.
 However, \eqn{sol23} does not have  any explicit invariance under
the  Galilean boosts 
\be\label{boos}
x^+\to x^+,~~~x^-\to x^--\vec v.\vec x +{v^2\over2}x^+,~~~\vec 
x\to~\vec x-\vec v x^+.
\ee
Thus the solution \eqn{sol23} represents a geometry with broken Lorentzian 
symmetry in which the time has  
dynamical exponent  $a=3$ and  which upon compactification to four 
dimensions simply gives us a 
Lifshitz  universe along with scalar and gauge fields. 
 The  background \eqn{sol23}
preserves at least 8  Poincar\'e supersymmetries. 

\subsection{Adding a boundary and resultant RG flow}
The classical solution \eqn{sol23} obtained under the vanishing horizon  
limits is devoid of an useful description near 
the boundary at infinity because $x^-$ tends to become null there while 
being a compact 
direction. However, there may be other possible ways to tackle this UV 
problem, 
in a rather adhoc manner we try to
add a boundary configuration to  the geometry \eqn{sol23}. So we 
  write  it down a little differently as 
\bea\label{sol23boun}
&&ds^2_{10}= r^2\left( -dx^{+}dx^{-}+{1\over 4}(1+{\beta^2\over  r^4}) 
(dx^{-})^2
+dx_1^2 +dx_2^2\right) +{dr^2\over  r^2}  + d\Omega_5^2  \br
&& F_5=4(1+*)Vol(S^5) 
\eea 
which  now includes a 
distinct UV configuration. We note, 
eventhough the solution \eqn{sol23boun} is obtainable from \eqn{sol23}
by incorporating a shift $x^{+}\to x^{+}-{1\over 4}x^{-}$, 
but with this shift the solution \eqn{sol23boun} has got a new asymptotics. 
Further it no longer has 
the invariance under asymmetric scalings in \eqn{a3}.  While
asymptotically (in the UV region) the  geometry \eqn{sol23boun}  becomes a 
pure anti-de Sitter spacetime 
\bea\label{sol23boun1}
&&ds^2_{10}= r^2\left( -dx^{+}dx^{-}+{1\over 4} 
(dx^{-})^2
+dx_1^2 +dx_2^2\right) +{dr^2\over  r^2}  + d\Omega_5^2  
\eea 
where the direction $x^-$  is no longer null.
This asymptotic pure AdS space has a flat boundary metric but obviously
in a `shifted' light-cone  basis and
has a relativistic scale invariance with dynamical exponent $a=1$ 
\be
r\to \xi^{-1} r , ~~~~~
x^{-}\to \xi^{1}  x^{-},~~~~
x^{+}\to \xi^1 x^{+},~~~x_{i}\to \xi x_{i}. 
\ee
 The dual conformal field theory in the shifted light-cone coordinates
will have a slightly  changed DLCQ description, 
see appendix. Particularly the lightcone energy 
spectrum will be different.  But the 
important thing to notice from \eqn{sol23boun} is that 
the size of the $x^-$ circle now remains finite in the UV region
including at the boundary. There may be some subtle issues involved in the 
CFT in doing this, as narrated above, but from 
gravity point of view we have got a classical background which is well 
behaved in UV and good
for holographic study. Also as discussed in the appendix, there 
appears to be no major difference in the two DLCQ descriptions so far as  
large lightcone momentum is involved. 
     
 In the deep IR region, as $r^4\ll \beta^2 $, the solution 
\eqn{sol23boun} 
becomes simply the Lifshitz solution 
 \eqn{sol23} we started with. Thus the  new type IIB background  
\eqn{sol23boun} 
describes a holographic flow of a
DLCQ theory (with $a=1$) in UV  to a nonrelativistic  Lifshitz theory 
(with $a=3$) in the IR.

\section{Vanishing horizon limits of TsT background}
Here we would like to focus on the
 `TsT' black D3-brane solutions which have Schr\"odinger like asymptotic 
symmetry. These 
solutions have $B$ field and can be 
obtained  from \eqn{sol2} by applying a chain of T-dualities and 
a `shift' \cite{malda}. The  TsT
solution in the string frame can be written as 
\bea\label{tst1}
&&ds^2=r^2\bigg[ H^{-1}\left(-{1+f\over 2}dx^{+}dx^{-}+{1-f\over 
4}[\lambda^{-2}
(dx^{+})^2 +\lambda^2 (dx^{-})^2 ]-
({\theta\over \lambda})^2 r^2 f (dx^+)^2\right)
\br && ~~~~~~~~~~ 
+dx_1^2 +dx_2^2\bigg] 
+{dr^2\over f r^2} +{\eta^2\over H } + ds^2({\cal B}_{KE}) \, \br
&& e^{2\phi}=1/H, ~~~~F_5=4 (1+\ast)Vol(SE)\br
&& B_{NS}={\theta\over 2\lambda} {r^2\over H}\big[(1+f)dx^{+} 
- (1-f)\lambda^2dx^{-}\big]\wedge\eta 
\eea 
where $H(r)=1+{\theta^2r_0^4\over r^2}$ and $\phi$ is the 
dilaton field. 
The  $\theta$ is 
the `shift' parameter  incorporated in the T-s-T duality \cite{malda} and 
it is  also 
related to the non-commutativity parameter  in light-like noncommutative 
description of gauge theories
\cite{ncym}; also see \cite{ncym1}. 
The $Vol(SE)$ is the volume form over the 
Sasaki-Einstein metric 
$$ds^2_{SE}=ds^2({\cal B}_{KE})+\eta^2 . $$ 
and 
$d\eta/2=J$ is the K\"ahler 2-form
over the base ${\cal B}_{KE}$ which is  K\"ahler-Einstein.
Unlike the `boosted' black D3-branes 
of the last section the  TsT black hole solutions \eqn{tst1} have 
desired Schr\"odinger  asymptotics at infinity which is
\bea\label{tst1a}
&&ds^2=r^2\big[-dx^{+}dx^{-}-\sigma^2r^2  (dx^+)^2
+dx_1^2 +dx_2^2 \big]
+{dr^2\over r^2} +{\eta^2 } + ds^2({\cal B}_{KE}) \, \br
&& e^{2\phi}=1, ~~~~F_5=4 (1+\ast)Vol(SE)\br
&& B_{NS}=\sigma r^2dx^{+} \wedge\eta, 
\eea 
where we  defined $\sigma=\theta/\lambda$. It has
 a scaling symmetry with dynamical exponent $a=2$, see 
\cite{malda} for further details. 

We are now interested in applying the vanishing horizon double limits 
\eqn{sol21} on the
TsT solutions \eqn{tst1}. In which case $H\sim1,~\sigma\sim 0$ and the 
resultant  zero temperature solution is
\bea\label{tst2aa}
&&ds^2=r^2\bigg[-dx^{+}dx^{-}+{\beta^2\over 
4r^4} (dx^{-})^2
+dx_1^2 +dx_2^2\bigg] 
+{dr^2\over  r^2} 
+{\eta^2 } + ds^2({\cal B}_{KE}) \, \br
&& e^{-2\phi}=1, ~~~~F_5=4(1+\ast)Vol(SE),~~~B_{NS}\sim0
\eea 
which is  a Lifshitz solution \eqn{sol23} discussed in the earlier section. 
That is Schroedinger asymptotics completely decouples from \eqn{tst2aa}
under the double scaling limits given in \eqn{sol21}.
There is however another distinct 
way of implementing these double limits  which is as follows.
 We shall now take the limits \eqn{sol21} but at the same time also scale 
the noncommutativity parameter $\theta$
such that  
$\sigma$ stays finite. In which case the zero 
temperature TsT solution is
\bea\label{tst2}
&&ds^2=r^2\bigg[ H^{-1}\left(-dx^{+}dx^{-}+{\beta^2\over 
4r^4} (dx^{-})^2-\sigma^2r^2  (dx^+)^2\right)
+dx_1^2 +dx_2^2\bigg] 
+{dr^2\over  r^2} 
\br && ~~~~~~~~~~ 
+{\eta^2\over H } + ds^2({\cal B}_{KE}) \, \br
&& e^{-2\phi}=H, ~~~~F_5=4(1+\ast)Vol(SE)\br
&& B_{NS}={\sigma} r^2H^{-1}\big[dx^{+} 
- {\beta^2\over 2r^4}dx^{-}\big]\wedge\eta 
\eea 
where $H(r)=1+{\sigma^2\beta^2\over r^2}$. 
Note that  this  zero 
temperature  solution \eqn{tst2} is  a Schr\"odinger solution 
asymptotically (in UV) due to the presence of nontrivial $B$ field, but 
it also has got a nonrelativistic deformation which becomes prominent
in the IR region.
The parameter $\beta$ is effectively a
measure of this IR deformation. 
Let us now investigate what happens in the deep IR region
where $r^2\ll\sigma^2\beta^2$.
In this region  
\be
e^{2\phi}\approx {r^2\over\sigma^2\beta^2},
\ee
so as the string coupling becomes weaker the solution \eqn{tst2} flows 
into 
\newpage
\bea\label{tst3}
&&ds^2=r^2\bigg[ {r^2\over\sigma^2\beta^2}\left(-dx^{+}dx^{-}+{\beta^2\over 
4r^4} (dx^{-})^2-\sigma^2r^2  (dx^+)^2\right)
+dx_1^2 +dx_2^2\bigg] 
\br && ~~~~~~~~~~ 
+{dr^2\over  r^2} +{r^2\over\sigma^2\beta^2}{\eta^2 } + ds^2({\cal 
B}_{KE}) \, \br
&&~~~~\approx
\left(-{r^6\over \beta^2}(dx^{+})^2
+r^2(dx_1^2 +dx_2^2)\right)+ {1\over 4\sigma^2} (\Xi)^2
 +{r^2\over\sigma^2\beta^2}{\eta^2 } + ds^2({\cal B}_{KE}) \, \br
&& B_{NS}\approx -{1\over 2\sigma}\Xi\wedge\eta \ .
\eea 
The above solution \eqn{tst3} is a nontrivial  configuration 
at zero temperature but
there is no scaling symmetry which survives 
in this deep IR region unless we scale
parameter $\sigma$. 
A non-relativistic scale (dilatation) transformation may be
written as
\bea\label{newscal}
&&r\to \xi^{-1} r , ~~~~
\sigma\to \xi^{-1} \sigma , ~~~~\br &&
x^{-}\to \xi^{-1}  x^{-},~~~~
x^{+}\to \xi^3 x^{+},~~~x_{1,2}\to \xi x_{1,2}.
\eea
So under this asymmetric scaling a solution like \eqn{tst3} with given $\sigma$ 
will tranform into another solution 
with a new $\sigma$ parameter. 
\footnote{However, let us specifically
note that all the thermodynamic properties 
of the TsT black holes do not at all depend upon this non-commutativity 
parameter \cite{malda}. This is due to manifest duality symmetry of the 
effective 8-dimensional string theory. }

In summary, 
from solutions \eqn{tst2aa} and \eqn{tst2} we  note that
these TsT
 spacetimes supported by nontrivial string  fields
would have a Lifshitz ($a=3$) or asymptotic Schroedinger symmetries ($a=2$)
 depending upon the presence of $B$ field in them.    
Particularly background \eqn{tst2} is a nontrivial solitonic configuration 
with $B$-field for which as we 
go from IR region to the UV region, it basically 
flows 
to  a Schr\"odinger spacetime (with  $a=2$). The corresponding holographic 
dual will be a nonrelativistic CFT at zero temperature with suitable
operater deformation giving rise to this flow. 
In the absence of any $B$ field the zero temperature limits provide a Lifshitz 
universe \eqn{tst2aa}.

Although the $x^-$ radius becomes constant  in \eqn{tst3}, 
but we should  be 
careful about of the size of the Hopf circle $\eta$  
in this region. In the deep IR region where  $$L^2 
{r^2\over\sigma^2\beta^2} \le 
l_s^2 ,$$  $L$ being the AdS radius,   we cannot trust the 
classical string geometry \eqn{tst3}  as the size of 
 fibre $\eta$ will
become sub-stringy there.

\section{Conclusions}
We  have reviewed  the vanishing horizon 
double limits $r_0\to0,~\lambda\to \infty$   
of  `boosted' black D3-branes having  a compact  lightcone direction.  
The zero temperature Lifshitz solutions
obtained as a result of taking these limits are not well 
defined in the UV, however by suitably
adding appropriate boundary configuration 
we have been able to construct a different  solution 
\eqn{sol23boun}. These boundary configurations correspond to  pure AdS 
 spacetime. The particle spectrum  has got
  shifted energies.
The  solution \eqn{sol23boun} describes a flow from  
AdS universe in UV to 
 $a=3$ Lifshitz universe in IR. We 
also study similar zero temperature limits for TsT black hole
solutions which involve  $B$-field. Resulting zero temperature solutions 
 describe an  RG flow from  Schr\"odinger spacetime ($a=2$) in 
UV to a nonrelativistic universe in the IR which has no apparent
 scaling symmetry. 
 The latter class of 
Galilean solutions  have an 
instability in the deep IR where the fibre direction over 
the K\"ahler base becomes sub-stringy. 
The main distinction between two types of solutions 
is the presence of $B$-field.

\vskip.5cm
\noindent{\it Acknowledgements:}\\
I am grateful to the AS-ICTP, Trieste for the  hospitality where 
this work got culminated and for the financial support under the
Associateship program. We are very much  thankful to the anonymous referee 
for the comments  which led to major corrections.

\appendix{
\section{Non-relativistic theories and DLCQ:}
Let us consider a  theory
with the following  $(d+2)$-dimensional flat spacetime  metric 
\be
ds^2=-dx^{+}dx^{-}+{1\over 4} (dx^{-})^2+ d\vec x^2
\ee
where $x^{\pm}=t\pm y$ are the lightcone coordinates. We shall denote 
the respective conjugate momenta by $p_{\pm}$, and the indices can 
be raised up by using the metric. Then the mass 
shell condition, $p_\mu p^\mu=0$, for a `massless' particle  will 
 become
\be
4p_{+}p_{-}-{\vec p}^2 + (p_+)^2=0
\ee
In these  coordinates if we identify $-2p_{-}=M$, $M$ being  actual 
rest mass, and the lightcone energy as $E_{+}\equiv -p_{+} $, then  we 
can  have
\be\label{disp}
E_{+}= \sqrt{{\vec p}^2+ M^2 }-M.
\ee
It can be easily seen that if we set $E_+=E-M$,
Eq.\eqn{disp} is precisely the standard
relativistic mass shell relation for a free 
particle in Minkowski space with 
mass $M$ and total energy $E$  in $(d+1)$ dimensions. So the 
lightcone energies are just shifted from $E$ to $E_+=E-M$. The 
relation \eqn{disp} remains 
valid even in the massless case $(p_-=0)$. 

The momentum $(-p_-)$ 
is a continuous variable so far. We can discretize it by compactifying
$x^-$ direction on a circle, ($x^-\sim x^- + 2\pi r^-$), so that
$-p_-$ will  
be quantized as ${N\over r^-}$ for ($N\ge 0$). Thus the spectrum is 
separated into 
discrete mass sectors characterised by $N$. This is known as discrete 
light cone quantization (DLCQ) of a 
relativistic  theory, see for  further discussion  \cite{malda} and 
references therein.  It is clear from the energy-mass relation 
\eqn{disp} that
a nonrelativistic limit in our DLCQ theory is achieved only when we focus on a 
given momentum sector with very large $N$, which means a large $M$. 
In the  limit of  very large $M$, 
  $\vec p^2 \ll M^2$, we get from \eqn{disp}
\be 
E_{+}\simeq {\vec p^2\over 2M} + \cdots 
 \ee
where the  $\cdots$ indicates  suppressed relativistic corrections 
like the
standard $M  ({\vec v^2\over c^2})^2$ etc. Thus in a large 
light-cone momentum sector the DLCQ theory is precisely a non-relativistic 
theory.

Notice the difference, the DLCQ of a relativistic theory in ordinary 
Minkowski metric 
$ds^2=-dx^+ dx^-+d\vec x^2$ gives a mass shell condition where lightcone 
energy is
\be\label{r4}
E_+=-p_+={\vec p^2\over (-4p_-)}
\ee
which looks like the energy of a non-relativistic particle of mass 
$M\sim-2p_{-}$ from start. However, in reality Eq.\eqn{r4} can be trusted as a 
nonrelativistic expression only in the large 
momentum sectors with $N\gg 0$, $i.e.$ when $\vec p^2\ll -p_-$. 
There is also a subtle issue here particularly involving
the massless modes ($p_-=0$), where the DLCQ needs to be treated 
differently, see \cite{hp}. 

We however see that the two expressions \eqn{disp} and \eqn{r4} coincide  in 
the  large 
momentum sectors which is of our immediate interest for the DLCQ in the 
gravity  where we have included backreaction due to the large lightcone 
momentum. This analysis shows that 
the large lightcone momentum sectors are indeed 
non-relativistic. 


\end{document}